\documentclass{amsart}

\newtheorem{theorem}{Theorem}[section]

\newtheorem{corollary}[theorem]{Corollary}
\newtheorem{proposition}[theorem]{Proposition}

\theoremstyle{definition}
\newtheorem{definition}[theorem]{Definition}

\theoremstyle{remark}
\newtheorem{remark}[theorem]{Remark}

\newcommand{\cM}{{\mathcal M}}

\newcommand{\cS}{{\mathcal S}}

\newcommand{\tM}{\widetilde{\mathcal{M}}}

\newcommand{\I}{1\!{\mathrm l}}
\newcommand{\Rn}{{\rm I\!R}} %Reals
\newcommand{\Nn}{{\rm I\!N}} %Naturals
\newcommand{\Cn}{{\setbox0=\hbox{
$\displaystyle\rm C$}\hbox{\hbox
to0pt{\kern0.6\wd0\vrule height0.9\ht0\hss}\box0}}} %Complex

\numberwithin{equation}{section}

\newcommand{\Tr}{\mathrm{Tr}}

\begin{document}

\title{Why are Orlicz spaces useful for  Statistical Physics?}

\author{W. A. Majewski}

\address{Institute of Theoretical Physics and Astrophysics, The Gdansk University, Wita Stwosza 57\\
Gdansk, 80-952, Poland and Unit for BMI, North-West-University, Potchefstroom, South Africa\\
E-mail: fizwam@univ.gda.pl}

\author{L. E. Labuschagne}

\address{DST-NRF CoE in Math. and Stat. Sci,\\ Unit for BMI, \\Internal Box 209, School of Comp., Stat. \& Math. Sci.\\, NWU, Pvt. Bag X6001, 2520 Potchefstroom, South
Africa\\
E-mail: Louis.Labuschagne@nwu.ac.za}

\begin{abstract}
We review a new formalism based on Orlicz spaces for the description of  large regular statistical systems. Our presentation includes both classical and quantum systems. The presented approach has the advantage that statistical mechanics is much better settled.
\end{abstract}

\maketitle

\noindent

\section{Introduction}
The basic mathematical ingredient of statistical physics, both classical and quantum, is a dual pair modeling the states and observables of the system under consideration:
\begin{enumerate}
\item
for classical physics
\begin{equation}
\label{1int}
( L^{\infty}(X, \mu), L^1(X, \mu)),
\end{equation}
where $(X, \mu)$ is a measure space, $L^1(X, \mu)= \{ f : \int_X |f|d\mu < \infty\}$, and where $L^{\infty}(X, \mu)$ stands for the essentially bounded, measurable functions on $X$. Here, we adopt the convenction, used in Physics, that the first component of the dual pair is related to observables; 
\item
for quantum physics:
\begin{equation}
\label{2int}
(B(\mathcal{H}), \mathfrak{F}_T(\mathcal{H})),
\end{equation}
where $\mathfrak{F}_T(\mathcal{H})$ denotes the trace class operators on a Hilbert space $\mathcal{H}$ while 
$B(\mathcal{H})$ stands for all linear bounded operators on $\mathcal{H}$.
\end{enumerate}

To support this claim, we remind that long ago, Maxwell had a remarkable idea, which was the germ of the second component
in (\ref{1int}). Making an analysis of ideal classical gases, he introduced the concept of a velocity distribution function $f$, i.e. a  function satisfying $f\geq 0$ and $\int f d\mu < \infty$. Then Boltzmann, elaborating 
the theory of ideal gases, obtained the so-called Boltzmann-equation, describing the time evolution of velocity distribution function.

The important point to note here is the following observation: it is common practice to interpret such a function $f$
as a mathematical device describing a state of a gas. Namely, in the simplest case, let us put $X = \Rn^3 \times \Rn$, and take $\mu$ to be the Lebesgue measure $dv$. Then $f(v,t) d^3v$ gives the density of particles in the volume element $d^3v$ centered at $v$, at the time $t$. Thus, as it was said, $f$ describes a state of a gas.

 Therefore, in that way, the considered convex subset of $L^1$-space, $\mathcal{S}_C = \{ f \in L^1(X, \mu); f \geq 0, \int_Xfd\mu = 1 \}$,  
has a nice physical interpretation. Note that $\mathcal{S}_C$ spans the $L^1$-space.
The first ingredient of the dual pair (\ref{1int}) is designed for a description of (bounded, classical) observables.

Turning to the quantum case, it is enough to note that the trace class operators $\mathfrak{F}_T(\mathcal{H})$ is the simplest example of non-commutative $L^1$-space, while $B(\mathcal{H})$ can be considered as an example of non-commutative $L^{\infty}$-space. Moreover, as in the previous case, density matrices describing states of a quantum system, $\mathcal{S}_Q = \{ \varrho \in \mathfrak{F}_T(\mathcal{H}); \varrho \geq 0, \Tr \varrho = 1 \}$,  form a convex generating subset of $\mathfrak{F}_T(\mathcal{H})$. Furthermore, self-adjoint elements of $B(\mathcal{H})$ describe (quantum) observables.

To sum up, in both cases, the dual pairs (\ref{1int}) and  (\ref{2int}) provide the starting point for statistical physics.

But, \textit{the crucial point to note here is the following observation}: for any $f \in L^{\infty}$ ($A \in B(\mathcal{H})$),
any $g \in \mathcal{S}_C$ ($\varrho \in \mathcal{S}_Q$ ) one has that, for any $n \in \Nn$, $\int f^n g d\mu < \infty$  ($\Tr \varrho A^n \varrho < \infty$ respectively!)
Consequently, in the standard approach to statistical physics described above, we are employing observables \textit{having all moments finite}. Thus if in more general settings we wish to have framework which preserves the essential character of the standard approach, then the above property of \textbf{finiteness of all moments}, should be taken as a rule for selecting an appropriate family of observables. We emphasize that applying such a rule suggests that an extension of the first components of (\ref{1int}) and (\ref{2int}) should be expected. To illustrate this, let us consider the quantum case. Then a well known result, see \cite{Win}, \cite{Wiel}, says \textit{it is impossible to realize canonical commutation relations in terms of a Banach algebra}. Consequently in looking for a framework in which such relations can be realized, it is natural to look for a larger family of observables than that given by $B(\mathcal{H})$.
Observables obtained in a manner that is faithful to the suggested procedure,  will be called regular observables and the corresponding system will be a regular system.  In other words,  we will be interested in the set consisting of all observables which have all moments finite.
It is worth pointing out that there is another line of reasoning in favour of such regular observables (see \emph{Introduction} in \cite{ML}).

A careful analysis of the structure of of classical regular observables led Pistone and Sempi, \cite{PS}, to the following result: such observables are described by the concrete Orlicz space, which is determined by the specific Young's function, $cosh - 1$.
But this means, among other things, that a description of regular observables demands a larger structure than $L^{\infty}$-space (see also the inclusions (\ref{Zygmunt})!). This has an important consequence. Namely, by duality
 the family of allowed states for a regular system will be smaller than the set of states given in terms of $L^1$-space only (again see inclusions (\ref{Zygmunt})).
Consequently, \textit{``a large portion'' of non-physical states will be removed} when passing to the set of allowed states for a regular system.

The aim of these notes, based on recent papers \cite{LM} and \cite{ML}, is to show how Orlicz spaces can be used for the improvement of traditional formalism used in statistical physics.
To this end, in Section 2, we review some of the standard facts on Orlicz spaces. Quantum Orlicz spaces are presented in Section 3. Section 4 (5) contains applications of Orlicz space technique to classical (quantum, respectively) regular statistical systems. Some conclusions are given in Section 6.

\section{ Classical Orlicz spaces}
 The classical $L^1(X, \Sigma, m)$, $L^2(X, \Sigma, m)$, $L^{\infty}(X, \Sigma, m)$
and the interpolating 
\newline
$L^p(X, \Sigma, m)$ spaces ($1\leq p < \infty$), where $(X, \Sigma, m)$ stands for a measure space,  may be regarded as spaces of measurable functions conditioned by the functions $t \mapsto |t|^p$ ($1\leq p < \infty$).
 The more general category of Orlicz spaces is defined as spaces of measurable functions conditioned by a more general class of convex functions; the so-called Young's functions. They are defined as

\begin{definition}
Let $\psi: [0, \infty) \to [0, \infty]$ be an increasing and left-continuous function such that $\psi(0)=0$. Suppose that on $(0,\infty)$ 
$\psi$ is neither identically zero nor identically infinite. Then the function $\Psi$ defined by
\begin{equation}
\label{psi}
\Psi(s) = \int_0^s \psi(u) du, \qquad (s\geq 0)
\end{equation}
is said to be a Young's function. We will assume that Young's functions are equal to $0$  for $x=0$.
\end{definition}

The functions:
 $x \mapsto |x|^p$, $x \mapsto \cosh(x) -1$, $x \mapsto x\log(x + \sqrt{1+x^2}) - \sqrt{1 + x^2} +1$, $x \mapsto xln(x +1)$ provide concrete examples of Young's functions.
To select a subclass of so called  regular Young's functions we will need:

\begin{definition}
\label{delta}
\begin{enumerate}
\item A Young's function $\Psi$ is said to satisfy the $\Delta_2$-condition if there	 exist $s_0 > 0$ and $c>0$ such that 
\begin{equation}
\Psi(2s) \leq c \Psi(s) < \infty, \qquad (s_0 \leq s < \infty).
\end{equation}
\item  A Young's function $\Phi$ is said to satisfy $\nabla_2$-condition if there exist $x_0 > 0$ and $l >1$ such that
\begin{equation}
\Phi(x) \leq \frac{1}{2l} \Phi(lx)
\end{equation}
for $x \geq x_0$.
\end{enumerate}
\end{definition}

In the theory of $L^p(X, \Sigma, m)$-spaces, the conjugate space $L^q(X, \Sigma, m)$, $\frac{1}{p} + \frac{1}{q} = 1$, in playing an important role. To have a generalization of this concept within the theory of Orlicz spaces we need:
\begin{definition}
 Let $\Psi$ be a Young's function, represented as in (\ref{psi}) as the integral of $\psi$. Define
 \begin{equation}
 \label{1}
 \phi(v) = \inf \{ w: \psi(w)\geq v \}, \qquad (0\leq v \leq \infty).
 \end{equation}
 Then the function
 \begin{equation}
 \label{2}
 \Phi(t) = \int_0^t \phi(v) dv, \qquad (0\leq t \leq \infty)
 \end{equation}
 is called the complementary Young's function of $\Psi$.
 \end{definition}

We note that if the function $\psi(w)$ is continuous and increasing monotonically then $\phi(v)$ is a function exactly inverse to $\psi(w)$. 
 Define (another Young's function)
\begin{equation}
x\log(x + \sqrt{1+x^2}) - \sqrt{1 + x^2} +1 = \int_0^x \mathrm{arcsinh}(v)dv.
\end{equation}

\begin{remark}
$x\log(x + \sqrt{1+x^2}) - \sqrt{1 + x^2} +1$  and $\cosh(x) -1$
are complementary Young's functions.
\end{remark}

Let $L^0$ be the space of measurable 
functions on some $\sigma$-finite measure space $(X, \Sigma, \mu)$. We will always assume,  that the considered measures
are $\sigma$-finite.
\begin{definition}
The Orlicz space 
$L^{\Psi}$ (being a Banach space) associated with  $\Psi$ is defined to be the set 
\begin{equation}\label{3}
L^{\Psi} \equiv L^{\Psi}(X, \Sigma, \mu) = \{f \in 
L^0 : \Psi(\lambda |f|) \in L^1 \quad \mbox{for some} \quad \lambda = \lambda(f) > 0\}.
\end{equation}
\end{definition}

 $L^{\Psi}$ can be equipped with two equivalent norms. The first one -
 Luxemburg-Nakano norm - is defined as
$$\|f\|_\Psi = \inf\{\lambda > 0 : \|\Psi(|f|/\lambda)\|_1 \leq 1\}.$$

while the second one -  Orlicz norm - for a pair $(\Psi, \Phi)$ of complementary Young's functions, is given by
$$\|f\|_\Phi = \sup\{ \int|fg|d\mu : \int\Psi(|g|) d\mu \leq 1 \}.$$

It is worth noting that $L^p$-spaces are nice examples of Orlicz spaces. 
The basic Orlicz spaces used in this paper are $L\log(L+1)$, $L^{\log}$ and $L^{\cosh - 1}$ defined by Young's functions: $x \mapsto x \log(x+1)$, $x \mapsto x\log(x + \sqrt{1+x^2}) - \sqrt{1 + x^2} +1$ and $x \mapsto \cosh(x) - 1$ respectively.
Other useful examples are provided by Zygmund spaces. They are defined as follows:
\begin{itemize}
\item $L\log L$ is defined by the following Young's function
$$s\log^+ s = \int_0^s \phi(u) du$$
where $\phi(u) =0 $ for $0\leq u\leq1$ and $\phi(u) = 1 + \log  u$ for $1< \infty $, where $\log^+x = \max (\log x, 0)$
\item $L_{\exp}$ is defined by the Young's function
$$ \Psi(s) = \int_0^s \psi(u)du, $$
where $\psi(0) = 0$ , $\psi(u) = 1$ for $0<u<1$, and $\psi(u)$ is equal to $e^{u -1}$ for $1 < u < \infty$.
Thus
$\Psi(s) = s$ for $0\leq s \leq 1$ and  $\Psi(s) = e^{s - 1}$ for $1< s < \infty$.
\end{itemize}

We recall that, for  a pair $(\Psi, \Phi)$ of complementary Young's functions with the function $\Psi$ satisfying $\Delta_2$-condition
there is the following relation  $(L^{\Psi})^* = L^{\Phi}$. In particular, $(L^{\log})^* = L^{\cosh - 1}$.

There is a natural question: what can be said about uniqueness of the correspondence: Young's function $\Psi \mapsto L^{\Psi}$-Orlicz space. To answer this question one needs the concept of equivalent Young's functions. To define it we will write $F_1 \succ F_2$ if and only if $F_1(bx) \geq F_2(x)$ for $x\geq 0$ and some $b>0$, and we say that the functions $F_1$ and $F_2$ are equivalent, $F_1 \approx
F_2$, if $F_1\prec F_2$ and $F_1\succ F_2$. One has (see \cite{RR})

\begin{theorem}
\label{2.6}
Let $\Phi_i$, $i =1,2$ be a pair of equivalent Young's function. Then $L^{\Phi_1} = L^{\Phi_2}$. 
\end{theorem}
Consequently, on condition that equivalence is preserved, \textit{one can ``manipulate'' Young's function's!} It is worth pointing out that the functions $x\log(x+1)$ and $ x\log(x + \sqrt{1+x^2}) - \sqrt{1 + x^2} +1 $ \textbf{are equivalent}.
Therefore, Theorem \ref{2.6} implies that $L^{\log} \equiv L\log(L+1)$. Moreover, Theorem \ref{2.6} also implies

\begin{proposition}
Let $(Y, \Sigma, \mu)$ be a $\sigma$-finite measure space and $L\log(L+1)$ be the Orlicz space defined by the Young's function $x \mapsto x\log(x+1), \ x\geq0$. Then $L\log(L+1)$ is an equivalent renorming of the K\"othe dual of $L^{\cosh -1}$.
\end{proposition}

For finite measure case (so, in particular, for a probability measure) Zygmund spaces appeared to be very useful. Namely

\begin{proposition}
\label{dwa}
For finite measure spaces $(X, \Sigma, m)$ one has
\begin{equation}
L^{\cosh - 1} = L_{\exp}.
\end{equation}
\end{proposition}
 
To understand the role of Zygmund spaces as well as why a description of all regular observables demands a larger structure than that given by $L^{\infty}$-space, the following result will be helpful, see \cite{BS}:
\begin{theorem}
\label{2.8}
Let $(Y, \Sigma, m)$ be a finite measure space with $m(Y) = 1$. The continuous embeddings
\begin{equation}
\label{Zygmunt}
L^{\infty} \hookrightarrow L_{\exp} \hookrightarrow L^p \hookrightarrow L\log L \hookrightarrow L^1
\end{equation}
hold for all p satisfying $1<p< \infty$. Moreover, $L_{exp}$ may be identified with the Banach space dual of $L\log L$.
\end{theorem}
It should be noted that for infinite measure case, the structure of inclusions is more complicated but they are in the same vein, see \cite{BS} for details.

To sum up, one has
\begin{corollary}
 The dual pair $(L^{\cosh -1}, L\log(L+1))$  will provide (cf Section 4) the basic mathematical ingredient for a description of a general, classical regular system while,
 for the finite measure case, the above pair of Orlicz spaces is an equivalent renorming of the pair of Zygmund spaces $( L_{\exp}, L\log L)$.
\end{corollary}

\section{Quantum Orlicz spaces}
To quantize the above outlined classical theory of Orlicz spaces, 
let $\Phi$ be a given Young's function. In the context of semifinite von Neumann algebras 
$\cM \subset B(\mathcal{H})$ equipped with an fns (faithful normal semifinite) trace $\tau$, the space of all $\tau$-measurable operators is defined as follows.
 Let $a$ be a densely defined closed operator on $\mathcal{H}$ with domain ${\mathcal D}(a)$ and let
$a = u|a|$ be its polar decomposition.
One says that $a$ is affiliated with $\cM$ (denoted $a \eta \cM$) if $u$ and all the spectral projections of $|a|$ belong to $\cM$.
 $a$ is $\tau$-measurable if $a \eta \cM$ and there is, for each $\delta >0$, a projection $e \in \cM$ such that $e\mathcal{H} \subset {\mathcal D}(a)$ and $\tau(1 - e)\leq \delta$.
We denote by $\widetilde{\cM}$ the set of all $\tau$-measurable operators.
$\widetilde{\cM}$ (equipped with the topology of convergence in measure) 
plays the role of 
$L^0$ (for details see \cite{nelson}, \cite{Ter}, and \cite{se}). In this  case, Kunze \cite{Kun} 
used this identification to define the associated noncommutative Orlicz space to be 
$$L^{ncO}_{\Phi}{} = \cup_{n=1}^\infty n\{f \in \widetilde{\cM} : \tau(\Phi(|f|) \leq 1\}$$ 
where $\Phi$ is a Young function, 
and showed that this is a linear space which becomes a Banach space when equipped with the 
Luxemburg-Nakano norm $$\|f\|_\Phi = \inf\{\lambda > 0 : \tau(\Phi(|f|/\lambda)) \leq 
1\}.$$
Moreover,
$$L^{ncO}_{\Phi}{} = \{f \in 
\widetilde{\cM} : \tau(\Phi(\lambda|f|)) < \infty  \quad \mbox{for some} \quad \lambda = 
\lambda(f) > 0\}.$$ 

However, to get a more tractable (and equivalent) quantization we will use the 
Dodds, Dodds, de Pagter approach \cite{DDdP} (see also Xu \cite{qxu}).
The first ingredient of this approach is the concept of generalized singular values. Namely,
given an element $f \in \widetilde{\cM}$ and $t \in [0, \infty)$, the generalized singular 
value $\mu_t(f)$ is defined by $\mu_t(f) = \inf\{s \geq 0 : \tau(\I - e_s(|f|)) \leq t\}$ 
where $e_s(|f|)$ $s \in \mathbb{R}$ is the spectral resolution of $|f|$. The function $t \to 
\mu_t(f)$ will generally be denoted by $\mu(f)$. For details on the generalized singular values 
see \cite{FK}.  Here, we note only that this directly extends classical notions where for any $f \in L^0{}$, 
the function $(0, \infty) \to [0, \infty] : t \to \mu_t(f)$ is known as the decreasing 
rearrangement of $f$. 

The second key ingredient of Dodds, Dodds, de Pagter approach is Banach Function Space. To define this concept, let $L^0(0, \infty)$
stands for
measurable functions on $(0, \infty)$ and $L^0_+$ denote $\{ f \in L^0(0,  \infty); f \geq 0\}$. A function norm 
$\rho$ on $L^0(0, \infty)$ is defined to be a mapping $\rho : L^0_+ \to [0, \infty]$ satisfying
\begin{itemize}
\item $\rho(f) = 0$ iff $f = 0$ a.e.  
\item $\rho(\lambda f) = \lambda\rho(f)$ for all $f \in L^0_+, \lambda > 0$.
\item $\rho(f + g) \leq \rho(f) + \rho(g)$ for all .
\item $f \leq g$ implies $\rho(f) \leq \rho(g)$ for all $f, g \in L^0_+$.
\end{itemize}
Such a $\rho$ may be extended to all of $L^0$ by setting $\rho(f) = \rho(|f|)$, in which case 
we may then define $L^{\rho}(0, \infty) = \{f \in L^0(0, \infty) : \rho(f) < \infty\}$. If 
now $L^{\rho}(0, \infty)$ turns out to be a Banach space when equipped with the norm 
$\rho(\cdot)$, we refer to it as a Banach Function space. If $\rho(f) \leq \lim\inf_n\rho(f_n)$ 
whenever $(f_n) \subset L^0$ converges almost everywhere to $f \in L^0$, we say that $\rho$ 
has the Fatou Property. If less generally this implication only holds for $(f_n) \cup \{f\} 
\subset L^{\rho}$, we say that $\rho$ is lower semi-continuous. If further the situation $f 
\in L^\rho$, $g \in L^0$ and $\mu_t(f) = \mu_t(g)$ for all $t > 0$, forces $g \in L^\rho$ and 
$\rho(g) = \rho(f)$, we call $L^{\rho}$ rearrangement invariant (or symmetric).

By employing  generalized singular values and Banach Function Spaces, 
 Dodds, Dodds and de Pagter \cite{DDdP} formally defined the noncommutative space 
$L^\rho(\widetilde{\cM})$ to be  $$L^\rho(\widetilde{\cM}) = \{f \in \widetilde{\cM} : \mu(f) \in 
L^{\rho}(0, \infty)\}$$ and showed that if $\rho$ is lower semicontinuous and $L^{\rho}(0, 
\infty)$ rearrangement-invariant, $L^\rho(\widetilde{\cM})$ is a Banach space when equipped 
with the norm $\|f\|_\rho = \rho(\mu(f))$. 

Now for any Young's function $\Phi$, the Orlicz 
space $L^\Phi(0, \infty)$ is known to be a rearrangement invariant Banach Function space 
with the norm having the Fatou Property, see Theorem 8.9 in \cite{BS}. Thus on selecting $\rho$ to be 
 $\|\cdot\|_\Phi$, the very general framework of Dodds, 
Dodds and de Pagter presents us with an alternative approach to realizing noncommutative 
Orlicz spaces.

We wish to close this section with the following concept, which will be used in the last section.
The space $L^\rho(\widetilde{\cM})$ is said to be fully symmetric if for any $f \in L^\rho(\widetilde{\cM})$ and $g \in \widetilde{\cM}$ the property $\int^{\alpha}_0 \mu_t(|g|)dt \leq
\int^{\alpha}_0 \mu_t(|f|) dt$, for any $\alpha >0$, ensures that $g \in L^\rho(\widetilde{\cM})$ with $\rho(g) \leq \rho(f)$.

\section{Applications of Orlicz spaces to Classical Statistical Mechanics}

To describe a classical regular system it is proposed to replace the dual pair

\begin{equation} 
 (L^{\infty}(X, \Sigma, m), L^1(X, \Sigma, m)),
 \end{equation}
by the following pair of Orlicz spaces (or equivalent pairs).
\begin{equation}
\label{Orlicz1}
 ( L^{\cosh -1}, \ L\log(L+1)),
 \end{equation}
where $L^{\cosh -1}$ and  $\ L\log(L+1)$ are classical Orlicz spaces described in Section 2.

To support this claim we observe:
\begin{enumerate}
\item Fix a measure space  $(X, \mu)$ and take $f \in \mathcal{S}_C$. Note that $f d\mu$ is a probability measure. Denote by $L(f\cdot \mu)$ the family of regular observables on $(X, f d\mu)$
which is defined in the following way: $L(f \cdot \mu)$ consists of all real random variables $u$ on $(X, \Sigma, fd\mu)$ such that $\hat{u}_f(t) = \int \exp(tu) f d\nu, \quad t \in \mathbb{R}$ \
is well defined in a neighborhood of the origin of $0$. The main objective of the above condition is to guarantee, see \cite{PS}, that all the moments of every $u \in L(f \cdot \mu)$ exist and they are the values at $0$ of the derivatives of $\hat{u}_f(t)$.
Pistone and Sempi proved \cite{PS}

\begin{theorem}
$L(f \cdot \mu)$ is the closed subspace of the Orlicz space
\newline
 $L^{\cosh - 1}(f\cdot \mu)$ of zero expectation random
variables.
\end{theorem}
\item  $H(f) = - \int f(x)\log f(x) d\mu$, $f \in {\cS}_{C}$ defines the classical continuous entropy. The principal significance of this function follows from the fact that it is strongly related to laws of thermodynamic.
However, for $f \in \cS_C$ the functional $H(f)$ is not well defined - see \cite{Bour}, Chapter IV, \S 6, Exercise 18.
\item But, if $f \in \cS_C^G = \{ f \in L\log(L+1), f\geq 0, \int f d\mu =1 \}$ then the situation is much improved. Namely, for example, in \cite{LM} one has the following 
\begin{proposition}\label{prop4.3}
Let $f \in L^1 \cap L \log(L+1)$  where both Orlicz spaces are over $(\Rn^3, \Sigma, d^3v)$ ($d^3v$ - the Lebesgue measure) and $f\geq 0$. Then
$$H_+(f) = \int f \log f d^3v$$
is bounded above, and if in addition $f\in L^{1/2}$ (equivalently $f^{1/2}\in L^1$), it is also bounded from below. Thus $H_+(f)$ is bounded below on a dense subset of the positive cone of $L \log(L+1)$.
\end{proposition}
\item The modern theory of Boltzmann equation, see \cite{DP1}, \cite{DP2}, \cite{AV}, and \cite{Vil} says that to build a mathematical theory of weak solutions of Boltzmann equation the condition
$f\in L \log(L+1)$ is indispensable. We remind that normalized, positive normalized  elements of the second component of (\ref{Orlicz1}) are describing states of a system. Thus, Boltzmann equation has its weak solutions in terms of regular states!
\item To sum up, the dual pair (\ref{Orlicz1}) has \textit{ the advantage of being general enough to encompass regular observables, and specific enough for the latter Orlicz space to select states with a well-defined entropy function and to describe weak solutions of Boltzmann equation}. 
\end{enumerate}
 
\section{Applications of Orlicz spaces to Quantum Statistical Mechanics}

To describe a quantum regular system, we fix a semifinite von Neumann algebra $\cM$ and a normal semifinite faithful trace $\tau$ acting on $\cM$. Then we propose to replace the dual pair

\begin{equation} 
 (B(\mathcal{H}), \mathfrak{F}_T(\mathcal{H})),
 \end{equation}
by the pair of noncommutative Orlicz spaces (or equivalent pairs)
\begin{equation}
\label{Orlicz2}
 ( L^{\cosh -1}, \ L\log(L+1)),
 \end{equation}
where now  $L^{\cosh -1}$ and  $\ L\log(L+1)$ are weighted versions of the non-commutative Orlicz spaces described in Section 3. More specifically, when considering how the Dodds, Dodds, de Pagter recipe (see Section 3) may be modified to obtain such weighted spaces, it makes sense to consider $L^{\cosh -1} \equiv L^{\cosh - 1}(\widetilde{\cM}) = \{f \in \widetilde{\cM} : \mu(f) \in 
L^{\cosh - 1}((0, \infty), \mu_t(x)dt)\}$ where $x \in \cM_*$, $x\geq 0$ ($\cM_*$ stands for the predual of $\cM$).  Obviously, here $\mu_t(x)$ is a non-commutative counterpart of the weight function $f$ appearing in the definition of classical weighted Orlicz space $L^{\cosh - 1}(f\cdot \mu)$. A similar definition gives a weighted counterpart of $L\log(L+1)$. It should be noted that the difference with the standard Dodds, Dodds, de Pagter approach, is that $L^{\rho}(0, \infty)$ has been replaced with $L^{\rho}((0, \infty), \mu(t)dt)$ in the definition of these spaces. Consequently, we are here concerned with objects that may validly be regarded as weighted noncomutative Orlicz spaces.

To support this claim we observe:
\begin{enumerate}
\item As for a classical case one can define the set of quantum regular observables $L^{quant}_x$. Namely (see \cite{LM}), 
\begin{definition}
\label{kwant}
\begin{equation}
L^{quant}_x = \{ g \in \tM: \quad 0 \in D(\widehat{\mu_{x}^g(t)})^0, \quad x \in m_{\tau}^+ \},
\end{equation}
where $D(\cdot)^0$ stands for the interior of the domain $D(\cdot)$ and
\begin{equation} 
\widehat{\mu_{x}^g(t)} = \int \exp(t\mu_s(g)) \mu_s(x)ds, \qquad t\in\mathbb{R}.
\end{equation}
Here, $\mu_s(g)$ stands for the generalized singular value of $g \in \tM$,
 $m_{\tau} = \{xy: x,y \in n_{\tau} \}$, where  $n_{\tau} = \{ x \in {\cM}: \tau(x^*x) < + \infty \}.$

We emphasize that the requirement that $0 \in D(\widehat{\mu_{x}^g(t)})^0$, presupposes that the transform $\widehat{\mu_{x}^g(t)}$ is well-defined in a neighborhood of the origin.
\end{definition}
\item As we are interested in a weighted version of the Dodds, Dodds, de Pagter approach, the well-definedness of such a version should be demonstrated. This is achieved by:
%%LL
\begin{theorem}[\cite{LM}]
\label{ncbf}
Let $x \in L_+^1(\cM, \tau)$. Let $\rho$ be a rearrangement-invariant Banach function norm on $L^0((0, \infty), \mu_t(x)dt)$ which satisfies the Fatou property and such that:
$\nu(E) < \infty \Rightarrow \rho(\chi) < \infty$ and $\nu (E) < \infty \Rightarrow \int_E f d\nu \leq C_E \rho(f)$
for some positive constant $C_E$, depending on $E$ and $\rho$ but independent of $f$ ($\nu$ stands for $\mu_t(x)dt$).
 Then $L^{\rho}_{x}(\tM)$ is a linear space and $\|\cdot\|_\rho$ a norm. Equipped with the norm $\|\cdot\|_\rho$, $L^{\rho}_{x}(\tM)$ is a Banach space which injects continuously into $\tM$.
\end{theorem}
\item Having the above generalization, the quantum counterpart of Pistone Sempi result is (see \cite{LM} and \cite{ML}):

\begin{theorem}[{(\cite{LM} \& \cite{ML})}]\label{QPS}
The set $L^{quant}_x$  coincides with the the weighted Orlicz space $L_x^{\cosh - 1}(\tM) \equiv L^{\Psi}_{x}(\tM)$ (where $\Psi = \cosh -1$) of noncommutative regular random variables.
\end{theorem}
Thus, again, quantum regular observables are described by specific Orlicz space; now by the quantum one.
\item Turning to quantum entropy, one has basically the same problems as for the classical one (see Section 4). But, again, changing $\cS_Q$ for positive normalized elements of the second component of the dual pair (\ref{Orlicz2}) the (quantum) entropy functional is more tractable. To illustrate this
we proved, see \cite{LM}

\begin{proposition}\label{prop6.8}
Let $\cM$ be a semifinite von Neumann algebra with an faithful normal semifinite trace $\tau$  and let $f \in L^1\cap L\log(L+1)(\tM)$, 
$f\geq 0$. Then
$\tau(f\log(f + \epsilon))$ is well defined for any $\epsilon >0$. Moreover
$$\tau(f \log f)$$
is bounded above, and if in addition $f\in L^{1/2}$ (equivalently $f^{1/2}\in L^1$), it is also bounded from below. Thus $\tau(f \log f)$ 
is bounded below on a dense subset of the positive cone of $L \log(L+1)$.
\end{proposition}

Consequently, again, the (quantum!) space $L\log(L+1)$ describes regular states.
\end{enumerate}

\begin{remark} It is worth noting that in the exceptional case of $B(\mathcal{H})$, the space $L^1\cap L\log(L+1)(B(\mathcal{H}))$ is exactly $\mathfrak{F}_T(\mathcal{H})$ (see the discussion at the end of \S 6 of \cite{ML}). Hence in the new approach suggested in these notes, we are not asking for a complete revision affecting even the standard approach, but rather for a proper extension of the standard approach when passing to systems having infinite degrees of freedom, i.e. to large systems.
Obviously, a system in statistical mechanics is par excellence such a system. Moreover (see \cite{Haag} and \cite{Thir}) $B(\mathcal{H})$ is not an adequate von Neumann algebra for a description of large systems. Consequently, the non-commutative integration theory based on $B(\mathcal{H})$ is not well suited for a description of Quantum Statistical Physics.
\end{remark}

\section{Conclusions}
We have presented strong arguments that the choice of the pair of Orlicz spaces $ ( L^{\cosh -1}, \ L\log(L+1))$ is better adapted for Statistical Physics, both classical as well as quantum,
than the traditional pair of Lebesgue $L^p$-spaces $(L^\infty, L^1)$.
In particular, the selection of regular states is well suited to handling with the entropy functional what is of crucial importance for thermodynamics.

But, as was recognized in early stage of Statistical Physics (Boltzmann equation) one needs to describe time evolution. This leads to a question of how this task can be carried out in terms of Orlicz spaces.
The first problem concerns possibility of lifting dynamical maps defined on the algebraic level, to maps given in terms of Orlicz spaces. This question was studied in \cite{LM}, and among other results, we proved:

\begin{theorem}\label{posmap}
Let $\cM_1, \cM_2$ be semifinite von Neumann algebras equipped with fns traces $\tau_1$ and $\tau_2$ respectively, and let $T: \cM_1 \to \cM_2$ be a positive map satisfying $\tau_2\circ T \leq C\tau_1$ for some constant $C > 0$. Then for 
any fully symmetric Banach function space $L^\rho(0, \infty)$, the restriction of \ $T$  to $\cM_1 \cap L^1(\cM_1, \tau_1)$ canonically extends to a bounded map from $L^\rho(\widetilde{\cM}_1)$ to $L^\rho(\widetilde{\cM}_2)$.
\end{theorem}
 
As Orlicz spaces (on $(0, \infty)$) are examples of fully symmetric Banach function spaces, the Dodds, Dodds, de Pagter approach described in Section 3, and Theorem  \ref{posmap} lead to well defined maps on, for example, quantum $L^{\cosh - 1}$-space (see \cite{LM} for details).
The natural subsequent question is related to Koopman's construction. We remind that this construction lies at the heart of ergodic theory. Again, an affirmative answer to this question can be provided. Namely, a large class of Jordan $*$-morphisms defined on a simifinite von Neumann algebra induce well defined composition operators on the (quantum) Orlicz space level, see \cite{LM} for details.

We wish to end these notes with the following challenging problem: How to describe dynamical semigroups in the framework of the considered pair of Orlicz spaces. In other words, we want to get a general description of dynamical systems for the new approach to Statistical Physics
and only in this way one can get a full-fledged theory. Our conjecture is that by applying interpolation techniques to the algebraic theory of dynamical maps, a complete description of a large class of dynamical semigroups will be obtained in terms of the considered Orlicz spaces.

\section{Acknowledgments}

The first-named author (W.A.M.) thanks for the support of the grant of the Foundation for Polish Science TEAM
project cofinanced by the EU European Regional Development Fund. The second-named author (L.E.L) wishes to declare that this work is based on research supported by the National Research Foundation, and that any opinion, findings and conclusions or recommendations expressed in this material, are those of the authors, and therefore that the NRF does not accept any liability in regard thereto.


\begin{thebibliography}{99}

\bibitem{AV} R. Alexandre, C. Villani, On the Boltzmann equation for long-range interactions, \textit{Comm. Pure Appl. Math.} \textbf{55} 30-70 (2002)
 
 
\bibitem{BS} G Bennet and R Sharpley, \textit{Interpolation of Operators}, 
Academic Press, London, 1988.
 
\bibitem{Bour} N. Bourbaki, {\em \'El\'ements de Math\'ematique. Livre VI: Int\'egration},  Hermann $\&$ $C^{ie}$ \'Editeurs, Paris, 1952

\bibitem{DP1} R. DiPerna, P.L. Lions, On the Fokker-Planck-Boltzmann equation, \textit{Commun. Math. Phys.} \textbf{120} 1-23 (1988) 

\bibitem{DP2} R. DiPerna, P.L. Lions, On the Cauchy problem for Boltzmann equations: Global existence and weak stability, \textit{Ann. Math} \textbf{130} 312-366 (1989)


\bibitem{DDdP} PG Dodds, T K.-Y Dodds and B de Pagter, Non-commutative Banach function spaces, \textit{Math Z} \textbf{201}(1989), 583-597.

\bibitem{FK} T Fack and H Kosaki, Generalized s-numbers of $\tau$- measurable operators, \textit{Pacific J Math} \textbf{123} (1986), 
269-300.

\bibitem{Haag} R. Haag, {\em Local Quantum Physics}, Springer Verlag, 1992

\bibitem{Kun} W Kunze, Non-commutative Orlicz spaces and generalized Arens algebras,
\textit{Math Nachr} \textbf{147}(1990), 123-138.


\bibitem{LM} L. E. Labuschagne, W. A. Majewski, Maps on non-commutative Orlicz spaces, \textit{Illinois J. Math}. {\bf 55}, 1053-1081, (2011)

\bibitem{ML} W. A. Majewski, L.E. Labuschagne, On applications of Orlicz spaces to Statistical Physics, \textit{ Ann. H. Poincare.}, {\bf 15}, 1197-1221, (2014)


\bibitem{nelson} E. Nelson, Notes on non-commutative integration, \textit{J. Funct. Anal.} \textbf{15} (1974), 103


\bibitem{PS} G. Pistone, C. Sempi, An infinite-dimensional geometric structure on the space of all the probability measures  equivalent to a given one, {\em Ann. Stat}. {\bf 23} (1995), 1543-1561 


\bibitem{se} I. E. Segal, A non-commutative extension of abstract integration, \textit{Ann. of Math.} \textbf{57} (1953), 401

\bibitem{RR} M. M. Rao, Z. D. Ren, {\em Theory of Orlicz spaces}, Dekker, 1991

\bibitem{Ter} M. Terp, {\it $L^p$ spaces associated with von Neumann algebras} Rapport No 3a (1981)

\bibitem{Thir} W. Thirring, \textit{A course in mathematical physics. Quantum Mechanics of large systems}, vol. 4, Springer, Berlin (1983)

\bibitem{Vil} C. Villani, A review of mathematical topics in collisional kinetic theory. in \textit{ Handbook of mathematical fluid dynamics, Vol. I}, North-Holland, Amsterdam,  2002, pp. 71-305
 
\bibitem{Win} A. Winter, The unboundedness of quantum mechanical matrices, {\textit{ Phys. Rev.}}, \textbf{71}, 737 -9 (1947)

\bibitem{Wiel} H. Wielandt,  \"Uder der unbeschr\"anktheit der operatoren der Quantum Mechanik, \textit{Math. Ann.} \textbf{121}, 21 (1949)

\bibitem{qxu} Q. Xu, Analytic functions with values in lattices and symmetric spaces of measurable operators,
\textit{Math. Proc. Camb. Phil Soc.} \textbf{109} (1991), 541--563.


\end{thebibliography}
\end{document}